\documentclass[epj]{svjour}

\usepackage{graphics}
\usepackage{amsmath,amssymb}

\begin{document}
\title{
  Analytical approach to the quantum-phase transition in the one-dimensional 
  spinless Holstein model
}
\author{S. Sykora\inst{1}, A. H\"{u}bsch\inst{2,3}, and K. W. Becker\inst{1}}
\institute{
  Institut f\"{u}r Theoretische Physik, Technische Universit\"{a}t Dresden, 
  D-01062 Dresden, Germany 
  \and
  Department of Physics, University of California, Davis, CA 95616, USA
  \and
  Max-Planck-Institut f\"{u}r Physik komplexer Systeme, N\"{o}thnitzer
  Stra{\ss}e 38, 01187 Dresden, Germany
}
\date{May 25, 2006}
\abstract{
  We study the one-dimensional Holstein model of spinless fermions interacting
  with dispersion-less phonons by using a recently developed projector-based
  renormalization method (PRM). At half-filling the system shows a 
  metal-insulator transition to a Peierls distorted 
  state at a critical electron-phonon coupling where both phases 
  are described within the same theoretical framework.
  The transition is accompanied by a phonon softening 
  at the Brillouin zone boundary and a gap in the electronic spectrum. 
  For different filling, the phonon softening appears away from the Brillouin
  zone boundary and thus reflects a different type of broken symmetry state.
  \PACS{
    {71.10.Fd}{Lattice fermion models (Hubbard model, etc.)}   \and
    {71.30.+h}{Metal-insulator transitions and other electronic transitions}
  }
}
\titlerunning{
  Quantum-phase transition in the Holstein model
}
\maketitle

\section{Introduction}

Systems with strong electron-phonon (EP) interactions have received
considerable attention in the last few years, motivated by new findings which
suggest a crucial role of the EP coupling 
in materials with strong electronic
correlations. Examples are high-temperature superconductors \cite{Lanzara},
colossal magnetoresistive manganites \cite{Millis}, or metallic alkaline-doped
C$_{60}$-based compounds \cite{Gunnarson}. Furthermore, in many quasi
one-dimensional materials, such as MX chains, conjugated polymers or organic
transfer complexes \cite{mat}, the kinetic energy of the electrons strongly
competes with the EP interaction which tends to establish, e.g.,
charge-density wave structures. 

The so-called spinless Holstein model,
\begin{eqnarray}
  \label{G1}
  {\cal H} &=&
  - t \sum_{\langle i,j\rangle} ( c_{i}^\dagger c_{j} + \mathrm{h.c.} )
  + \omega_0 \sum_i  \; b_i^\dagger b_i \\
  &&
  + g \sum_i \; (b_i^\dagger + b_i)n_i\nonumber,
\end{eqnarray}
is perhaps the simplest realization of a strongly coupled EP system. It
describes the local interaction $g$ at a given lattice site $i$
between the density $n_{i} = c^{\dagger}_{i}c_{i}$ of electrons and 
dispersion-less phonons with frequency  $\omega_0$. Here, the $c^{\dagger}_{i}$
($b^{\dagger}_{i}$) are fermionic (bosonic) creation operators of electrons
(phonons), and $\langle i,j\rangle$ denotes the summation over all neighboring
lattice sites $i$ and $j$. 

In the past, a large number of different analytical and numerical methods have
been applied to the Holstein mo\-del \eqref{G1}. 
Strong-coupling
expansions \cite{HF83}, variational \cite{ZFA89} and renormalization
group \cite{CB84,BGL95} approaches, as well as Monte Carlo \cite{HF83,MHM96} 
simulations were used 
to investigate mainly ground-state properties. More
recently, exact diagonalization \cite{WF98} (ED) 
and density matrix
renormalization group (DMRG) \cite{BMH98,jecki,FWH04} techniques, and dynamical
mean-field theory (DMFT) in conjunction with a numerical renormalization group
approach \cite{Meyer} were applied. However, most of these approaches are
restricted in their applications, e.g., ED techniques can only handle very
small system sizes, DMRG methods require one-dimensional systems, and the DMFT
exploits the limit of infinite spatial dimensions. Furthermore, the phononic
part of the Hilbert space is infinite even for finite systems so that 
all numerical approaches require truncation schemes to limit the
number of bosonic degrees of freedom, or a numerically expensive systematic
reduction of the Hilbert space in the spirit of the DMRG method \cite{Zhang}
has to be employed. For these reasons, there is still a clear need of reliable
theoretical methods to tackle strongly coupled EP systems in terms of minimal
theoretical models. The Holstein model of spinless fermions \eqref{G1} shows
at half-filling a quantum-phase transition from a metallic to an insulating
state where both the one-dimensional case and the limit of infinite dimensions
have been studied \cite{HF83,BMH98,Meyer}.

\bigskip
Alternative analytical approaches to interacting many-particle systems are
offered by methods based on functional renormalization like the flow equation
method \cite{Wegner}, the similarity transformation \cite{Wilson}, or the
projector-based renormalization method (PRM) \cite{Becker}. So far, these
methods have been successfully applied to electron-phonon systems with the aim
to study the effective electron-electron interaction \cite{Lenz} or
superconductivity \cite{Hubsch_SC}. However, the quan\-tum-phase transition in
the Holstein model has not yet been studied by this kind of approach.

Recently, we applied the projector-based renormalization method (PRM)
\cite{Becker} to the spinless Holstein model \eqref{G1} at half-filling
\cite{Sykora}. This {\it analytical} approach does not suffer from the
truncation of the phononic Hilbert space so that {\it all bosonic} degrees of
freedom are taken into account. Furthermore, the PRM treatment provides both
fermionic and bosonic quasi-particle energies which are \textit{not} directly
accessible with other methods. However, the PRM approach of Ref.~\cite{Sykora}
was restricted to the metallic phase.

Here, we extend our recent work \cite{Sykora} on the one-dimen\-sional spinless
Holstein model \eqref{G1} where the scope of this paper is twofold: Firstly,
we demonstrate that the PRM approach of Ref.~\cite{Sykora} is not restricted
to the half-filled case, and that the phonon dispersion relation close to the
critical value of the EP coupling reflects the type of the broken symmetry of
the insulating phase. Secondly, we derive modified renormalization equations
for the half-filled case that enable a dimerization of the system. Thus, we
find an {\it uniform} description for the metallic as well as for the
insulating phase of the spinless Holstein model \eqref{G1} at half-filling. 
We determine the critical coupling $g_{c}$ for the metal insulator transition
where a careful finite-size scaling is performed. Furthermore, a phonon
softening is found for EP couplings close to the critical value $g_{c}$ which
can be understood as a precursor effect of the phase transition. 

\bigskip
This paper is organized as follows. In the next section we briefly describe
the basic idea of our recently developed PRM approach \cite{Becker}, and
discuss metallic solutions of the renormalization equations for different
fillings of the electronic band. In particular, we show that the phonon
softening close to the critical value $g_{c}$ of the metal insulator
transition reflects the type of the broken symmetry of the insulating
phase. In section \ref{uniform}, we extend the PRM approach of
Ref.~\cite{Sykora} to derive an uniform description for the metallic as well
as for the insulating phase of the spinless Holstein model \eqref{G1} at
half-filling. Furthermore, in section \ref{results}, the critical coupling 
$g_{c}$ for the metal insulator transition is determined, and electronic as
well as phononic quasi-particle energies are presented. Finally, we summarize
in section \ref{summary}.

\section{Methodology and metallic solutions}
\label{metal}

In the PRM approach \cite{Becker}, the final effective Hamiltonian 
$\tilde{\mathcal{H}} = \lim_{\lambda\rightarrow 0} \mathcal{H}_{\lambda}$ is
obtained by a sequence of unitary transformations,
\begin{eqnarray}
  \label{G2}
  \mathcal{H}_{(\lambda-\Delta\lambda)} &=& 
  e^{X_{\lambda,\Delta\lambda}} \, \mathcal{H}_{\lambda} \,
  e^{-X_{\lambda,\Delta\lambda}},
\end{eqnarray}
by which transitions between eigenstates of the unperturbed part 
$\mathcal{H}_{0}$ of the Hamiltonian caused by the interaction
$\mathcal{H}_{1}$ are eliminated in steps. The respective transition energies
are used as renormalization parameter $\lambda$. The generator 
$X_{\lambda,\Delta\lambda}$ of the unitary transformation has to be adjusted
in such a way so that $\mathcal{H}_{(\lambda-\Delta\lambda)}$ does only
contains excitations with energies smaller or equal
$(\lambda-\Delta\lambda)$. Interaction with energies larger than
$(\lambda-\Delta\lambda)$ are used up to renormalize the parameters of the
effective Hamiltonian. Thus, difference equations for the $\lambda$ dependence
of the parameters of the Hamiltonian can be derived which we call
renormalization equations.

\bigskip
In Ref.~\cite{Sykora} we have made the following ansatz
\begin{eqnarray}
  \label{new_G3}
  \mathcal{H}_{0,\lambda} &=& 
  \sum_{k} \varepsilon_{k,\lambda} c^{\dagger}_{k}c_{k} + 
  \sum_{q} \omega_{q,\lambda} b^{\dagger}_{q} b_{q} + E_{\lambda}, \\[1ex]
  \label{new_G4}
  \mathcal{H}_{1,\lambda} &=&
  \frac{1}{\sqrt{N}}\sum_{k,q} g_{k,q,\lambda} 
  \left(
    b^{\dagger}_{q} c^{\dagger}_{k}c_{k+q} + 
    b_{q} c^{\dagger}_{k+q}c_{k}
  \right)
\end{eqnarray}
for the renormalized Hamiltonian 
$\mathcal{H}_{\lambda}=\mathcal{H}_{0,\lambda} + \mathcal{H}_{1,\lambda}$
after all excitations with energies larger than $\lambda$ have been
eliminated. In the next step, all excitations within the energy shell between
$(\lambda-\Delta\lambda)$ and $\lambda$ have been removed where 
\begin{eqnarray*}
  X_{\lambda,\Delta\lambda} &=&
  \frac{1}{\sqrt{N}}\sum_{k,q} B_{k,q,\lambda,\Delta\lambda}
  \left(
    b^{\dagger}_{q} c^{\dagger}_{k}c_{k+q} -
    b_{q} c^{\dagger}_{k+q}c_{k}
  \right)  
\end{eqnarray*}
has been used as generator of the unitary transformation \eqref{G2}. The
coefficients $B_{k,q,\lambda,\Delta\lambda}$ have to be fixed in such a way so
that only excitations with energies smaller than $(\lambda-\Delta\lambda)$
contribute to $\mathcal{H}_{(\lambda-\Delta\lambda)}$.

Evaluating Eq.~\eqref{G2}, operator terms that contain four fermionic and
bosonic one-particle operators appear. However, to restrict the
renormalization scheme to the operators of the renormalization ansatz
according Eqs.~\eqref{new_G3} and \eqref{new_G4} we have to perform a
factorization approximation,
\begin{eqnarray*}
  c^{\dagger}_{k}c_{k} c^{\dagger}_{k-q}c_{k-q} &\approx&
  c^{\dagger}_{k}c_{k} \langle c^{\dagger}_{k-q}c_{k-q} \rangle + 
  \langle c^{\dagger}_{k}c_{k} \rangle c^{\dagger}_{k-q}c_{k-q} \\
  && -
  \langle c^{\dagger}_{k}c_{k} \rangle 
  \langle c^{\dagger}_{k-q}c_{k-q} \rangle, \\[1ex]
  b^{\dagger}_{q} b_{q} c^{\dagger}_{k}c_{k} &\approx& 
  b^{\dagger}_{q} b_{q} \langle c^{\dagger}_{k}c_{k} \rangle + 
  \langle b^{\dagger}_{q} b_{q} \rangle c^{\dagger}_{k}c_{k}  -
  \langle b^{\dagger}_{q} b_{q} \rangle
  \langle c^{\dagger}_{k}c_{k} \rangle .
\end{eqnarray*}
Thus, the renormalization equations for $\varepsilon_{k,\lambda}$,
$\omega_{q,\lambda}$, $E_{\lambda}$, and $g_{k,q,\lambda}$, that are obtained
by comparing the result of the transformation \eqref{G2} with
Eqs.~\eqref{new_G3} and \eqref{new_G4}, depend on unknown expectation values 
$\langle c^{\dagger}_{k}c_{k} \rangle$ and 
$\langle b^{\dagger}_{q} b_{q} \rangle$. These expectation values are best
defined with $\mathcal{H}_{\lambda}$ since the renormalization step is done
from $\mathcal{H}_{\lambda}$ to $\mathcal{H}_{(\lambda-\Delta\lambda)}$.  Note
that for simplicity in Ref.~\cite{Sykora} the expectation values were evaluated
 with the unperturbed Hamiltonian $\mathcal{H}_{0,\lambda}$. In contrast, we
 now neglect the $\lambda$ dependence of the expectation values and perform
 the factorization approximation with the full Hamiltonian $\mathcal{H}$ 
in order to take into account important interaction effects. (A general
discussion of the factorization approximation in the PRM can be found in
Ref.~\cite{Hubsch_PAM}.) 

For the evaluation of the expectation values with the full Hamiltonian
$\mathcal{H}$ we have to apply the unitary transformations also on operators
to exploit 
$
\langle{\cal A}\rangle = \langle {\cal A}_\lambda \rangle_{{\cal H}_\lambda} 
$. 
The transformed operator ${\cal A}_{\lambda}$ is obtained 
by the sequence \eqref{G2} of unitary transformations,
${\cal A}_{(\lambda-\Delta \lambda)}= e^{X_{\lambda,\Delta \lambda}} 
{\cal A}_\lambda e^{-X_{\lambda, \Delta \lambda}}$.
We derive renormalization equations for the fermionic and bosonic
one-particle operators, $c_{k}^{\dag}$ and $b_{q}^{\dag}$, where
the same approximations are used as for the Hamiltonian. In this way, 
equations for the needed expectation values are obtained.

The resulting set of renormalization equations has to be solved
self-consistently. The explicit (numerical)
evaluation starts from the cutoff $\lambda=\Lambda$ 
of the original model and proceeds down to $\lambda=0$. 
Note that the case $\lambda=0$ with self-consistently determined expectation
values provides the effectively free model 
$\tilde{\mathcal{H}}=\lim_{\lambda\rightarrow 0}\mathcal{H}_{0,\lambda}$,
\begin{eqnarray}
  \label{new_G5}
  \tilde{\mathcal{H}} &=& 
  \sum_{k} \tilde{\varepsilon}_{k}
  c^{\dagger}_{k}c_{k} + 
  \sum_{q} \tilde{\omega}_{q}
  b^{\dagger}_{q} b_{q} + 
  \tilde{E},
\end{eqnarray}
we are interested in. Here, we have introduced the renormalized dispersion
relations 
$\tilde{\varepsilon}_{k}=\lim_{\lambda\rightarrow 0}\varepsilon_{k,\lambda}$
and   
$\tilde{\omega}_{q}=\lim_{\lambda\rightarrow 0}\omega_{q,\lambda}$, and the 
energy shift $\tilde{E}=\lim_{\lambda\rightarrow 0}E_{\lambda}$. 

\begin{figure}
\begin{center}
  \scalebox{0.65}{
    \includegraphics*{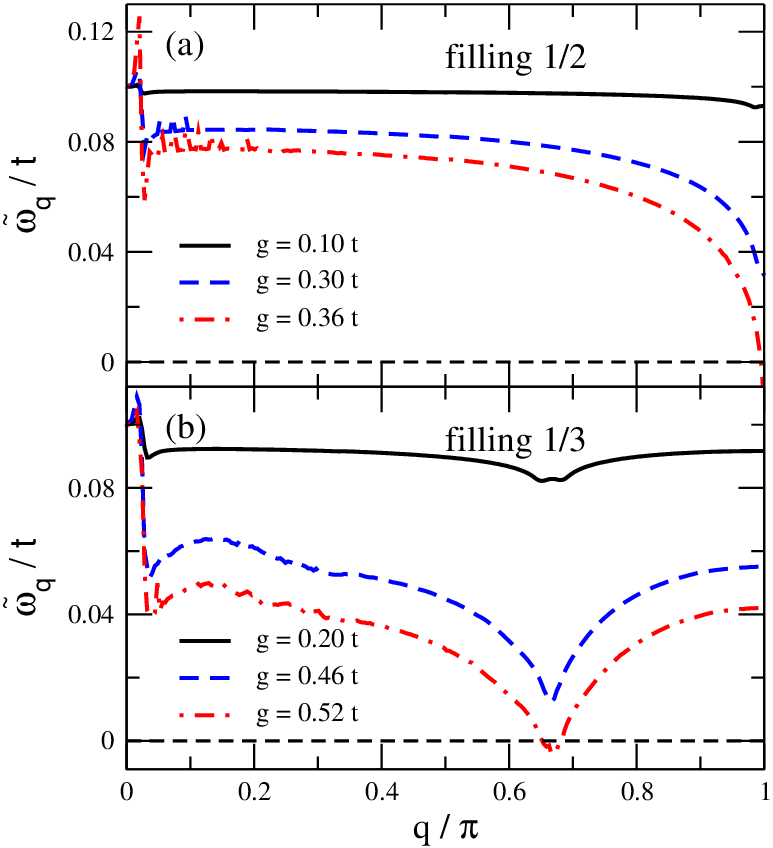}
  }
\end{center}
\caption{
  (color online) Bosonic quasi-particle energies of a chain with 500 lattice
  sites for different EP couplings $g$ obtained from the renormalization
  equations without a symmetry breaking term. Note the unphysical jumps 
  at small q values which are due to the factorization approximation of higher
  order renormalization processes. For details see the discussion in
  Ref.~\cite{Sykora}. 
} 
\label{Fig_1}
\end{figure}

\bigskip
Here, one might wonder why it is possible to map the Holstein model of spinless
fermions onto an effectively \textit{free} system as described above. Of
course, before the actual calculations can be started a guess of the form of
the final renormalized Hamiltonian $\tilde{\cal H}$ is needed that can be
motivated by the operator terms generated due to the renormalization
procedure. Furthermore, the obtained results are a very powerful test of the
inital guess: Unphysical findings are clear signatures of an insufficient
renormalization ansatz. As an example, it will turn out that non-physical
negative phonon energies $\tilde\omega_{q}$ are obtained if the
electron-phonon coupling $g$ exceeds a critical value. Therefore, we shall
later extend ansatz \eqref{new_G3}, \eqref{new_G4} to study both the metallic
and the insulating phase of model \eqref{G1}. 

Furthermore, it is important to note that the employed factorization
approximation is directly related to the re\-nor\-ma\-li\-za\-tion ansatz: Only
operator structures that are contained in the renormalization ansatz
$\mathcal{H}_{\lambda}$ can be taken into account without any
approximation. All other operators appearing due to the renormalization
procedure have to be traced back to those of $\mathcal{H}_{\lambda}$. Here, as
already mentioned, we use a factorization approximation for this purpose. Due
to the complexity of the renormalization equations it is extremely difficult to
estimate the effect of the mentioned factorization approximation on the
results. However, in this paper we present two complementary renormalization
schemes so that the comparison of the two approaches will provide some
information about the effects of the employed factorization approximations
(see the discussion in Sec.~\ref{results}).

\begin{figure}
\begin{center}
  \scalebox{0.65}{
    \includegraphics*{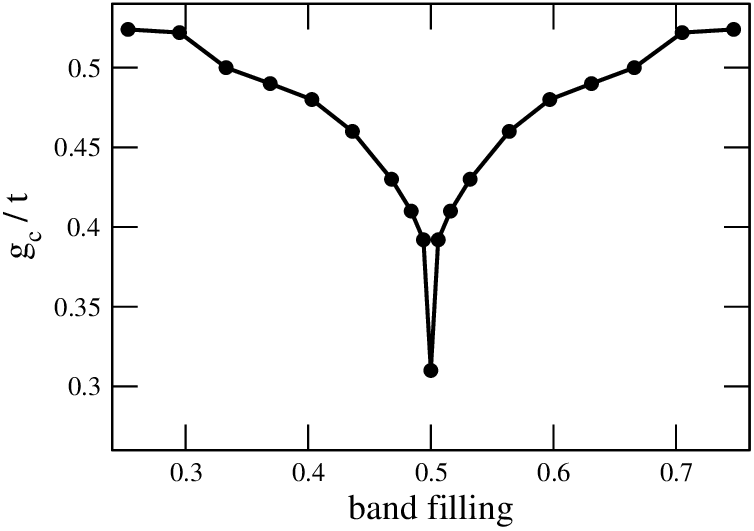}
  }
\end{center}
\caption{
  Critical value $g_{c}$ of the EP coupling as a function of the electronic
  band filling for a chain with 500 lattice sites. The value 
  $g_{c}\approx 0.31t$ for the half-filled case agrees perfectly with the
  result of Ref.~\cite{Sykora} where a simpler scheme for the evaluation of
  the expectation values has been used.
}
\label{Fig_2}
\end{figure}

\bigskip
In the following we concentrate on the so-called adiabatic case 
$\omega_{0}\ll t$, where we have chosen $\omega_{0}/t = 0.1$. 
In panel (a) of Fig.~\ref{Fig_1}, the bosonic dispersion relation of an
one-dimensional chain at half-filling is shown for different EP couplings
$g$. Due to the coupling between the phononic and electronic degrees of
freedom, $\tilde{\omega}_{q}$ gains some dispersion. In particular, a
phonon softening appears at the Brillouin-zone boundary if $g$ is only
slightly smaller than the critical value $g_{c}\approx 0.31t$ of the EP
coupling. However, if the EP coupling $g$ exceeds the critical
value $g_{c}$ we obtain non-physical negative phonon energies. Therefore, the
solutions of the renormalization equations from Ref.~\cite{Sykora} can only
describe the metallic phase and break down at the phase transition.

As already mentioned above, in the present calculation an improved scheme for
the evaluation of the expectation values has been used. However, the same
critical value $g_{c}\approx 0.31t$ for the metal-insulator transition was
also found in Ref.~\cite{Sykora} where all expectation values had been
evaluated using the unperturbed part $\mathcal{H}_{0,\lambda}$ of the
$\lambda$ dependent Hamiltonian and not using the full Hamiltonian
$\mathcal{H}$.

Note that the critical EP coupling $g_{c}\approx 0.31t$ obtained from the
vanishing phonon phonon mode at the Brillouin-zone boundary is significantly
larger than the DMRG value of $g_c \approx 0.28 t$
\cite{BMH98,FWH04}. This difference can be understood as follows: Our
renormalization scheme as presented in this Sec. starts from a description
particularly suitable for the metallic phase as represented by the final
Hamiltonian $\tilde{\mathcal{H}}$. As discussed above, a factorization
approximation has been employed so that fluctuations are suppressed. However,
these additional fluctuations would enhance the pho\-non softening and,
therefore, tend to destabilize the metallic phase. Thus, the realistic 
critical value $g_{c}$ of the electron-phonon coupling is smaller than the
result we have obtained by the PRM approach as presented above. Note that the
situation will change for the modified renormalization scheme of
Sec.~\ref{uniform}. 

\bigskip
Even though only the case of half-filling has been discussed in
Ref.~\cite{Sykora}, the same PRM approach can also be applied to different
fillings without any modifications. Panel (b) of Fig.~\ref{Fig_1} shows
bosonic dispersion relations for an one-dimensional chain at filling 1/3 for
different EP couplings. Similar to the half-filled case, we obtain a pho\-non
softening for EP couplings $g$ close to a critical value of
$g_{c}\approx 0.5t$, and negative phonon energies are observed for
$g>g_{c}$. However, in contrast to the half-filled case, the phonon softening
appears at $2k_F=2\pi/3$ and not at the Brillouin-zone boundary (where $k_{F}$
denotes the Fermi momentum). Consequently, as expected for a one-dimensional
fermionic system, the broken symmetry of the insulating phase {\it strongly}
depends on the filling of the electronic band. 

Fig.~\ref{Fig_2} shows the critical values $g_{c}$ of the EP coupling as 
function of the electronic band filling. Here, two aspects of the results are
important to be noticed: (i) The critical value $g_{c}$ has a minimum at
half-filling which is connected with the highest stability of the insulating
phase. (ii) The form of the curve in Fig.~\ref{Fig_2} reflects the
particle-hole symmetry of the model \eqref{G1}.

\bigskip
Above we have argued that the phonon softening is a precursor effect of the
metal-insulation transition in the Holstein model. However, in the
anti-adiabatic limit of the model where $t\ll\omega_{0}$ holds we do obtain
the opposite behavior: A phonon stiffening occurs in the metallic phase if we
increase the electron-phonon coupling $g$. This finding corresponds to the
anti-adiabatic limit of spin systems where also no phonon softening occurs for
fast phonons \cite{spin}. Unfortunately, in this work the metal-insulator
transition itself can not be studied in the anti-adiabatic limit because the
presented PRM approach has no stable solutions for $t\ll g$.

\section{Uniform description of metallic and insulating phases at half-filling}
\label{uniform}

In this section we want to extend our PRM approach to find an uniform
description for the metallic as well as for the insulating 
phase of the spinless Holstein model \eqref{G1}. In the following, we
concentrate on the case of half-filling because the broken
symmetry of the insulating phase depends strongly on the filling of the
electronic band as discussed above.

\bigskip
We have already mentioned that the PRM approach as discussed in
Sec.~\ref{metal} breaks down for strong EP couplings. In this case a
long-range charge density wave order occurs, and electrons and ions are
shifted from their symmetric positions. To describe such a state with a broken
symmetry in the framework of the PRM the ansatz for $\mathcal{H}_{\lambda}$
must contain suited order parameters \cite{Hubsch_SC}. In the case of the
half-filled Holstein model, one has to take into account that the unit cell of
the system is doubled in the case of a dimerized insulating
ground-state. Therefore, we consider the Hamiltonian in a reduced Brillouin
zone and introduce appropriate symmetry breaking terms in our renormalization
ansatz. Thus, the renormalized Hamiltonian 
$\mathcal{H}_{\lambda} = \mathcal{H}_{0,\lambda} + \mathcal{H}_{1,\lambda}$ 
reads
\begin{eqnarray}
  \label{G3}
  \lefteqn{\mathcal{H}_{0,\lambda} \,=\,} \\
  & = & 
  \sum_{k>0,\alpha}
  \varepsilon_{\alpha,k,\lambda}
  c_{\alpha,k}^{\dag} c_{\alpha,k} + 
  \sum_{q>0, \gamma} 
  \omega_{\gamma,q,\lambda}
  b_{\gamma,q}^{\dag} b_{\gamma,q} + 
  E_{\lambda} 
  \nonumber \\
  &&
  + \sum_{k}
  \Delta_{k,\lambda}^{\mathrm{c}} 
  \left( 
    c_{0,k}^{\dag} c_{1,k} + \mathrm{h.c.} 
  \right) 
  + \sqrt{N}
  \Delta_{\lambda}^{b} 
  \left( b_{1,Q}^{\dag} + \mathrm{h.c.} \right)
  \nonumber\\[2ex]
  \label{G4}
  \lefteqn{\mathcal{H}_{1,\lambda} \,=\,} && \nonumber\\
  & = &
  \frac{1}{\sqrt{N}} 
  \sum_{
    \genfrac{}{}{0pt}{1}{
      \genfrac{}{}{0pt}{1}{k, q>0}{
        \alpha,\beta,\gamma
      }
    }{}
  }
  g_{k,q,\lambda}^{\alpha,\beta,\gamma}
  \left\{ \delta(b_{\gamma,q}^{\dag})
  \delta(c_{\alpha,k}^{\dag}
  c_{\beta,k+q}) + \mathrm{h.c.} \right\}
\end{eqnarray}
after all excitations with energies larger than $\lambda$ have been integrated
out. Here, due to the usage of the reduced Brillouin zone, both the fermionic
and bosonic one-particle operators as well as the model parameters have
additional band indices, $\alpha,\beta,\gamma = 0, 1$. Furthermore, the 
definitions $\delta A = A - \langle A \rangle$ and $Q=\pi/a$ are used 
in Eq.~\eqref{G4}. Note, however, that Eq.~\eqref{G3} is restricted to the
one-dimensional case at half-filling but could be easily extended: For higher
dimensions the term with the order parameter $\Delta_{\lambda}^{\mathrm{b}}$
has to take into account all $\mathbf{Q}$ values of the Brillouin zone
boundary. To describe the insulating phase away from half-filling one has to
adjust the order parameters because the broken symmetry type strongly depends
on the filling as discussed above.

\bigskip
To derive the renormalization equations we consider the renormalization step
from $\lambda$ to $(\lambda-\Delta \lambda)$. At first we perform a
rotation in the fermionic subspace and a translation to new ionic equilibrium
positions so that $\mathcal{H}_{0,\lambda}$ is diagonalized,
\begin{eqnarray}
  \label{G5}
  \mathcal{H}_{0,\lambda} &=& 
  \sum_{k>0} \sum_{\alpha} 
  \varepsilon_{\alpha,k,\lambda}^{C}
  C_{\alpha,k,\lambda}^{\dag} C_{\alpha,k,\lambda} \\
  &&
  + \sum_{q>0}
  \sum_{\gamma} \omega_{\gamma,q,\lambda}^{B} 
  B_{\gamma,q,\lambda}^{\dag}
  B_{\gamma,q,\lambda} - E_{\lambda}.
  \nonumber
\end{eqnarray}
Next we rewrite $\mathcal{H}_{1,\lambda}$ in terms of the new fermionic
and bosonic creation and annihilation 
operators, $C_{\alpha,k,\lambda}^{(\dag)}$ and
$B_{\gamma,q,\lambda}^{(\dag)}$ which we have introduced to diagonalize
$\mathcal{H}_{0,\lambda}$. Finally, we have to evaluate \eqref{G2} 
to derive the renormalization equations for the parameters of
$\mathcal{H}_{\lambda}$.  Here the ansatz 
\begin{eqnarray*}
  X_{\lambda, \Delta \lambda}
  &=&
  \frac{1}{\sqrt N} \sum_{k,q} \sum_{\alpha,\beta,\gamma}
  A_{k,q,\lambda,\Delta \lambda}^{\alpha,\beta,\gamma} \\
  &&
  \quad \times
  \left\{
    \delta B_{\gamma,q}^\dagger 
    \delta(
      C_{k,\lambda}^\dagger C_{\beta,k+q,\lambda}
    ) 
    - \mathrm{h.c.}
  \right\}
\end{eqnarray*}
is used for the generator of the unitary transformation \eqref{G2}. The
coefficients  $A_{k,q,\lambda,\Delta \lambda}^{\alpha,\beta,\gamma}$ have to
be fixed in such a way so that only excitations with energies smaller than
$(\lambda-\Delta\lambda)$  contribute to 
$\mathcal{H}_{1,(\lambda- \Delta \lambda)}$. The renormalization equations 
for the parameters $\epsilon_{k,\lambda}, \Delta^c_{k,\lambda},
\omega_{\gamma, q, \lambda}, \Delta^b_\lambda$, and $g^{\alpha, \beta,
  \gamma}_{k, q, \lambda}$ are obtained by comparison with \eqref{G3},
\eqref{G4} after the creation and annihilation operators
$C^{(\dagger)}_{\alpha,k,\lambda},B^{(\dagger)}_{\gamma, q,\lambda}$ 
have been transformed back to the original operators
$c^{(\dagger)}_{\alpha,k}, b^{(\dagger)}_{\gamma, q}$. The actual calculations
are done in close analogy to Ref.~\cite{Sykora}. 

\bigskip
As already discussed in Sec.~\ref{metal}, in order to evaluate Eq.~\eqref{G2},
an additional factorization approximation must be employed where only
operators of the structure of those of \eqref{G3} and \eqref{G4} are
kept. At this point it is important to notice that this factorization
approximation is performed in the framework of the reduced Brillouin
zone. Therefore, in comparison to the PRM scheme of Sec.~\ref{metal},
additional terms occur for non-zero order parameters 
$\Delta_{k,\lambda}^{c}$ and $\Delta_{\lambda}^{b}$. 
Due to the factorization approximation, 
the final renormalization equations still depend on unknown
expectation values. As already discussed in Sec.~\ref{metal}, these
expectation values were evaluated with the full Hamiltonian $\mathcal{H}$ in
order to take into account important interaction effects. Therefore, we again
must apply the sequence \eqref{G2} of unitary transformations to operators, 
$
  {\cal A}_{(\lambda-\Delta \lambda)}= e^{X_{\lambda,\Delta \lambda}} 
  {\cal A}_\lambda e^{-X_{\lambda, \Delta \lambda}}
$
to exploit
$
\langle{\cal A}\rangle = \langle {\cal A}_\lambda \rangle_{{\cal H}_\lambda} 
$.
As in Sec.~\ref{metal}, this procedure is performed for the fermionic
and bosonic one-particle operators, $c_{\alpha,k}^{\dag}$ and
$b_{\gamma,q}^{\dag}$, where the same approximations are used as for the
Hamiltonian. Thus, we easily obtain equations for the needed expectation
values. The resulting set of renormalization equations is solved numerically
where the equations for the expectation values are taken into account due to a
self-consistency loop.

\bigskip
The case $\lambda=0$ with self-consistently determined expectation values
provides again an effectively free model  
$
  \tilde{\mathcal{H}} = \lim_{\lambda\rightarrow 0} \mathcal{H}_{\lambda}
  = \lim_{\lambda\rightarrow 0} \mathcal{H}_{0,\lambda}
$ which reads 
\begin{eqnarray}
\label{G6}
  \tilde{ \mathcal{H}} & =&  \sum_{k>0,\alpha} 
\tilde{\varepsilon}_{\alpha,k}
   c_{\alpha,k}^{\dag} c_{\alpha,k} + 
   \sum_{k>0}
   \tilde{\Delta}_{k}^{\mathrm{c}} 
   \left( 
     c_{0,k}^{\dag} c_{1,k} + \mathrm{h.c.} 
   \right) \\
   &+& \sum_{q>0, \gamma}
\tilde{\omega}_{\gamma,q}
   b_{\gamma,q}^{\dag} b_{\gamma,q} + \sqrt{N}
   \tilde{\Delta}^{b} 
   \left( b_{1,Q}^{\dag} + b_{1,Q} \right)
  - \tilde{E} \nonumber
\end{eqnarray}
where  
$
\tilde{\varepsilon}_{\alpha, k} = 
\lim_{\lambda \rightarrow 0}{\varepsilon}_{\alpha, k,\lambda}
$, 
$
  \tilde{\Delta}_{k}^{c} = 
  \lim_{\lambda\rightarrow 0}\Delta_{k,\lambda}^{c}
$,
$
\tilde{\omega}_{\gamma, q}=
\lim_{\lambda \rightarrow 0} \omega_{\gamma, q, \lambda}
$,
and
$
  \tilde{\Delta}^{b} = 
  \lim_{\lambda\rightarrow 0}\Delta_{\lambda}^{b}
$.
Note that all excitations from ${\cal H}_{1,\lambda}$ were used up to
renormalize the parameters of $\tilde{\cal H}_0$. 
The expectation values are also calculated in 
the limit $\lambda\rightarrow 0$ and can be easily determined from  
$
  \langle A \rangle_{\mathcal{H}} 
  = \langle A_{\lambda} \rangle_{\mathcal{H}_{\lambda}}
  = \langle 
      (\lim_{\lambda\rightarrow 0} A_{\lambda}) 
    \rangle_{\tilde{\mathcal{H}}}
$
because $\tilde{\mathcal{H}}$ is a free model.

\bigskip
Before we present results in Sec.~\ref{results}, we want to compare the two
different renormalization schemes discussed above in more detail. As already
mentioned, the approach of Sec.~\ref{metal} is based on a renormalization
ansatz in particular suitable for the metallic phase of the system. In
contrast, the renormalization scheme as discussed in the present section starts
from a dimerized system so that this approach provides a description of the
system particularly adapted to the insulating phase. Thus, both
renormalization schemes are complementary approaches, and opposite
fluctuations are suppressed due to the employed factorization approximations:
The suppressed fluctuations of the renormalization scheme of Sec.~\ref{metal}
\textit{destabilize} the metallic phase. In contrast, it will be shown that
the neglected fluctuations of the approach of Sec.~\ref{uniform} \textit{favor}
the same phase. 

On the other hand, the two renormalization schemes are not only complementary
in their starting points but also closely related: The renormalization scheme
as discussed in the present section \textit{exactly} matches the approach of
Sec.~\ref{metal} if the gap parameters $\Delta^{c}_{k,\lambda}$ and 
$\Delta^{b}_{\lambda}$ vanish for all $\lambda$ values. Thus, finite
excitation gaps are only obtained from the renormalization scheme of this
section if the insulating (i.e. gapped) phase has a lower free energy than the
metallic solution as discussed in Sec.~\ref{metal}. Note that in actual
calculations the PRM approach of this section \textit{always} leads to a free
energy smaller than the one obtained from the scheme discussed in
Sec.~\ref{metal}.

\begin{figure}
\begin{center}
  \scalebox{0.65}{
    \includegraphics*{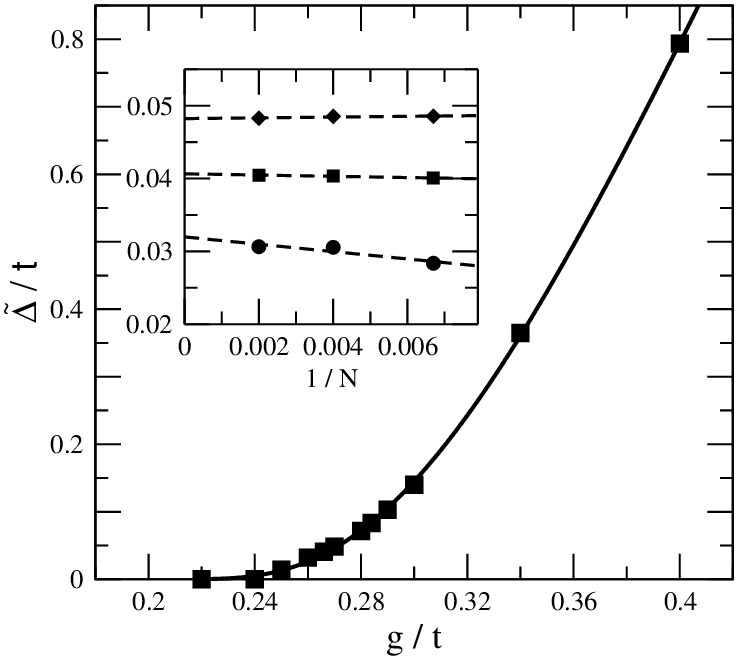}
  }
\end{center}
\caption{
  Gap in the electronic excitation spectrum for the infinite chain. The solid
  line is a Kosterlitz-Thouless fit of the form 
  $
    \tilde{\Delta}/t = \frac{15.240}{\sqrt{(g/t)^{2} - (0.195)^{2}}} \; 
    \mathrm{exp}\left[-\frac{1.400}{\sqrt{(g/t)^{2} - (0.195)^{2}}}\right]
  $. 
  The inset shows the finite-size scaling for $g$ values of $0.26 t$
  (circles), $0.266 t$ (squares), and $0.27 t$ (diamonds).
}
\label{Fig_3}
\end{figure}

\section{Results}
\label{results}

In the following, we first show that the PRM treatment can be used to
investigate the Peierls transition of the one-dimensional spinless Holstein
model \eqref{G1} at half-filling. In particular, our analytical approach
provides a theoretical description for the metallic as well as for the
insulating phase, and we compare our results with DMRG calculations
\cite{BMH98,FWH04}. 

\bigskip
In order to determine $g_{c}$ we perform a careful finite-size scaling as
shown for some $g$ values in the inset of Fig.~\ref{Fig_3} where a linear
regression is applied to extrapolate our results to infinite system size. Note,
however, that the finite size scaling of our PRM approach is affected by two
different effects: Long-range fluctuations are suppressed by the finite
cluster size as well as by the used factorization approximation so that a
rather unusual dependence on the system size is found. In Fig.~\ref{Fig_3} the
electronic excitation gap $\tilde{\Delta}$ for infinite system size,
as found from the opening of a gap in the quasi-particle energy
$\tilde{\varepsilon}_k$ (see text below), is plotted as function of the EP
coupling $g$. A closer inspection of the data shows that an insulating phase
with a finite excitation gap is obtained for $g$ values larger than the
critical EP coupling $g_{c}\approx 0.24t$. A comparison with the critical
value $g_{c}\approx 0.28t$ obtained from DMRG calculations \cite{BMH98,FWH04}
shows that the critical values from the PRM approach of Sec.~\ref{uniform}
might be somewhat too small.

In particular, the critical coupling $g_{c}\approx 0.24t$ obtained from the
opening of the gap in $\tilde{\varepsilon}_k$ is significantly smaller than
the $g_{c}$ value of $\approx 0.31t$ which was found from the vanishing of the
phonon mode at the Brillouin zone boundary in the metallic solution of
Sect.~\ref{metal}. Instead, one would expect that both the gap in
$\tilde{\varepsilon}_{k}$ and the vanishing of $\tilde{\omega}_{q}$ should
occur at the same $g_c$ value. This inconsistency can be understood from the
different nature of the factorization approximations employed in the two PRM
approaches presented here: As discussed above, the due to the factorization
approximation suppressed fluctuations tend to destabilize the suggested phase
of the renormalization ansatz. Therefore, the neglected fluctuations of the two
presented PRM approaches have opposite characters: The stability of the
metallic phase is overestimated by the scheme of Sec.~\ref{metal} whereas the
approach of Sec.~\ref{uniform} favors the insulating phase, and the different
values of the critical coupling exactly reflect the different nature of the
two PRM approaches. Thus, both approaches (and both ways to determine $g_c$)
would be consistent with each other if additional fluctuation could be taken
into account, and one would expect a common result for $g_c$ between $0.24 t$
and $0.31 t$. This would be in good agreement with the DMRG value of $g_c
\approx 0.28 t$ \cite{BMH98,FWH04}.

Another way to determine $g_{c}$ might be given by the assumption that a
marginal relevant interaction in a renormalization group treatment causes the
phase transition in the Holstein model. 
Corresponding to Ref.~\cite{White}, the gap $\tilde{\Delta}$ 
should in this case grow for $g>g_{c}$ as $\exp[-a/g-g_{c}]$, where $a$ is some
numerical constant. Such a function fits our data very well, and a critical
value of $g_{c}=0.166t$ would result in this way which seems to be
questionable small. On the other hand, we could also assume that the
metal-insulator transition might be of Kosterlitz-Thouless type
\cite{Kosterlitz} as found for the anti-adiabatic limit of spin-Peilerls
chains \cite{Caron}. As one can see in Fig.~\ref{Fig_3}, the
Kosterlitz-Thouless gap formula \cite{Baxter,Caron},
$
  \tilde{\Delta} \propto (g^{2} - g_{c}^{2})^{-0.5} \,
  \mathrm{exp}(-a / \sqrt{g^{2} - g_{c}^{2}})
$,
also fits our data and leads to a critical value of $g_{c}=0.195t$. However, 
such a small value contradicts other finding of our calculations: As discussed
above, the phonon softening at the Brillouin-zone boundary is a clear
signature of the occuring phase transition, and the smallest phonon energies
should be obtained for $g_{c}$. This test provides clear evidence for a
critical value of $g_{c}\approx 0.24t$.

\begin{figure}
  \begin{center}
    \scalebox{0.65}{
      \includegraphics*{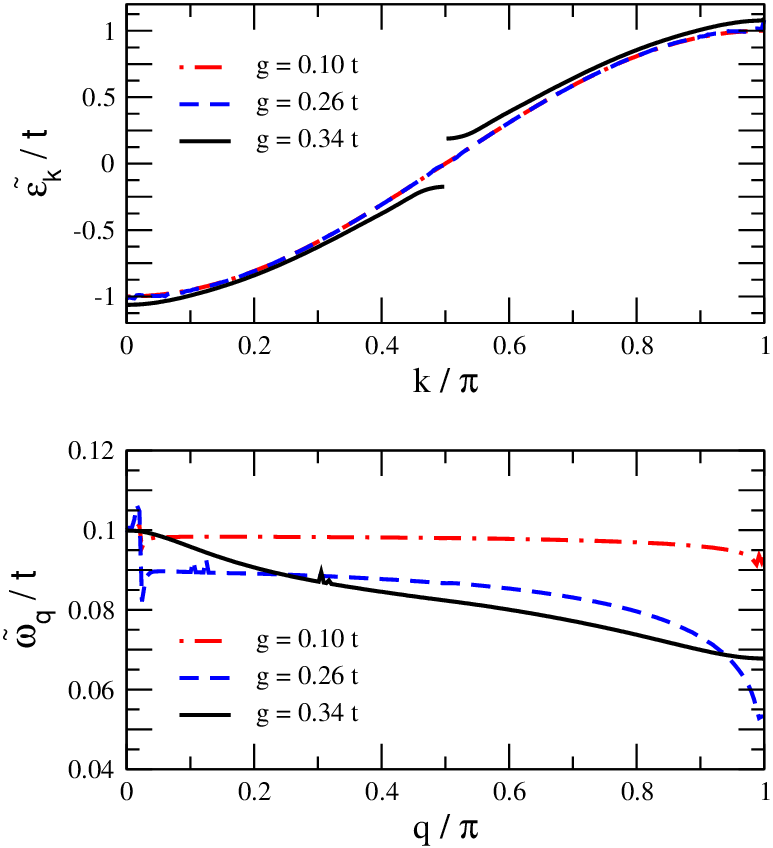}
    }
  \end{center}
  \caption{
    (color online) Fermionic (upper panel) and bosonic (lower panel)
    quasi-particle energies of a chain with $500$ lattice sites for different
    EP couplings $g$.
  }
  \label{Fig_4}
\end{figure}

\bigskip
The quasi-particle energies are also directly accessible:
After the renormalization equations were
solved self-consistently the electronic and phononic quasi-particle energies
of the system, $\tilde{\varepsilon}_{k}$ and 
$\tilde{\omega}_{q}$, respectively, are given by the limit 
$\lambda\rightarrow 0$ of the parameters 
$\varepsilon_{\alpha,k,\lambda}^{C}$ and 
$\omega_{\gamma,q,\lambda}^{B}$ of the unperturbed part
$\mathcal{H}_{0,\lambda}$ of the Hamiltonian in its diagonalized form
\eqref{G5}. 

In Fig.~\ref{Fig_4} the renormalized one-particle energies (which have to be
interpreted as the quasi-particle energies of the full system) 
are shown for different values of
the EP coupling $g$. As one can see from the upper panel, the electronic
one-particle energies depend only slightly on $g$ as long as $g$ is smaller
than the critical value $g_{c}\approx 0.24t$. If the EP coupling $g$ is
further increased a gap $\tilde{\Delta}$ opens at the Fermi energy so that the
system becomes an insulator. Remember 
that the gap $\tilde{\Delta}$ has been used
as order parameter to determine the critical EP coupling $g_{c}$ of the
metal-insulator transition (see Fig.~\ref{Fig_3}). The lower panel of
Fig.~\ref{Fig_4} shows the results for the phononic one-particle energies 
$\tilde{\omega}_{q}$. 
One can see that $\tilde{\omega}_{q}$ gains dispersion 
due to the coupling $g$ between the electronic and phononic
degrees of freedom. In particular, the phonon mode at momentum $2k_F=\pi$,
i.e.~at the Brillouin-zone boundary, becomes soft if the EP coupling is
increased up to 
$g_{c}\approx 0.24t$. However, in contrast to the metallic solution of
Sect.~\ref{metal} $\tilde{\omega}_q$ at $2k_F$ always remains positive 
though it is very small. Note that for $g$ values larger than $g_c$
$\tilde{\omega}_q$ increases again.
This phonon softening at the phase transition has to be
interpreted as a lattice instability which leads to the formation of the
insulating Peierls state for $g>g_{c}$. The phase transition is
associated with a shift of the ionic equilibrium positions. A lattice
stiffening occurs if $g$ is further increased to values 
much larger than the critical value  $g_{c}\approx 0.24t$.

\section{Summary}
\label{summary}

In this paper, the recently developed PRM approach has been applied to the
one-dimensional Holstein model \eqref{G1} of spinless fermions interacting with
dispersion-less phonons. In extension to our earlier work \cite{Sykora},
here, we have improved the scheme for the evaluation of expectation values,
and have discussed the metal-insulator transition of the model for different
fillings of the electronic band. Furthermore, for the
half-filled Holstein model  we have derived an uniform
description for the metallic as well as for the insulating phase. 

We have shown that the renormalized phonon energies gain a momentum dependence
due to the EP coupling. In particular, for half-filling 
a phonon softening at $2 k_F$ appears if the EP coupling is close to the
critical value of the 
metal-insulator transition. Therefore, the broken symmetry of the insulating
phase strongly depends on the filling of the electronic band. The critical
value of the metal-insulator transition also depends on the band filling, 
where the insulating phase has at half-filling the highest stability.

The PRM approach of Ref.~\cite{Sykora} breaks down if the EP coupling
exceeds the critical value of the metal-insulator transition. Here, we have
extended our PRM approach to enable symmetry broken insulating states. Thus, we
derived an \textit{uniform} description for the metal as well as for the
insulating phase (which is restricted to the half-filled case of the Holstein
model). Here, we have used this extended analytical framework to study the
phase transition in more detail. We have determined the critical value of the
EP coupling for the metal-insulator transition. Furthermore, we have shown the
opening of the excitation gap in the electronic quasi-particle energies, and
the lattice stiffening for EP couplings larger than the critical value.

\bigskip
\textit{Acknowledgment.}
We would like to acknowledge helpful discussions with T.M. Bryant. 
This work was supported by the DFG through the research program SFB 463 and
under Grant No. HU~993/1-1, and by the US Department of Energy, Division of
Materials Research, Office of Basic Energy Science.



\begin{thebibliography}{}
\bibitem{Lanzara} A.~Lanzara {\it et al.}, Nature {\bf 412}, 510 (2001).
\bibitem{Millis} A.J.~Millis, P.B.~Littlewood, and B.I.~Shraiman,
  Phys. Rev. Lett.~{\bf 74}, 5144 (1995).
\bibitem{Gunnarson} O.~Gunnarson, Rev. Mod. Phys. {\bf 69}, 575 (1997).
\bibitem{mat} A.R.~Bishop and B.I.~Swanson, Los Alamos Science
  {\bf 21}, 133 (1993); N.~Tsuda, K.~Nasu, A.~Yanese, K.~Siratori, 
  {\it Electronic Conduction in Oxides} (Springer-Verlag, Berlin, 1990);
  J.-P.~Farges (Ed.), {\it Organic Conductors} (Marcel Dekker, New York 1994).
\bibitem{HF83} J.E.~Hirsch and E.~Fradkin, Phys.~Rev.~B {\bf 27}, 4302 (1983).
\bibitem{ZFA89} H.~Zheng, D.~Feinberg, and M.~Avignon, Phys.~Rev.~B {\bf 39},
  9405 (1989). 
\bibitem{CB84} L.G.~Caron and C.~Bourbonnais, Phys.~Rev.~B {\bf 29}, 4230
  (1984). 
\bibitem{BGL95} G.~Benfatto, G.~Gallovotti, and J.L.~Lebowitz,
  Helv. Phys. Acta {\bf 68}, 312 (1995). 
\bibitem{MHM96} R.H.~McKenzie, C.J.~Hamer, and D.W.~Murray, Phys. Rev. B
  {\bf 53}, 9676 (1996).
\bibitem{WF98} A.~Wei{\ss}e and H.~Fehske, Phys. Rev. B {\bf 58}, 13526
  (1998); H.~Fehske, M.~Holicki, and A.~Wei{\ss}e, Advances in Solid State
  Physics {\bf 40}, 235 (2000). 
\bibitem{BMH98} R.J.~Bursill, R.H.~McKenzie, and C.J.~Hamer,
  Phys. Rev. Lett. {\bf 80}, 5607 (1998).
\bibitem{jecki} E.~Jeckelmann, C.~Zhang, and S.R.~White, Phys. Rev. B 60,
  7950-7955 (1999). 
\bibitem{FWH04} H.~Fehske, G.~Wellein, G.~Hager, A.~Wei\ss e, K.W.~Becker,
  and A.R.~Bishop, Physica B {\bf 359-361}, 699 (2005). 
\bibitem{Meyer} D.~Meyer, A.C.~Hewson, and R.~Bulla, Phys. Rev. Lett. 
  {\bf 89}, 196401 (2002).
\bibitem{Zhang} C.~Zhang, E.~Jeckelmann, and S.R.~White, Phys. Rev. B {\bf 80},
  2661 (1998).
\bibitem{Wegner} F.J.~Wegner, Ann. Physik (Leipzig) {\bf 3}, 77 (1994).
\bibitem{Wilson} S.D.~G{\l}azek and K.G. Wilson, Phys. Rev. D {\bf 48}, 5863
  (1993); \textit{ibid.} {\bf 49}, 4214 (1994).
\bibitem{Becker} K.W.~Becker, A.~H\"{u}bsch, and T.~Sommer, Phys. Rev. B
  {\bf 66}, 235115 (2002).
\bibitem{Lenz} P.~Lenz and F.~Wegner, Nucl. Phys. B {\bf 482}, 693 (1996).
\bibitem{Hubsch_SC} A.~H\"{u}bsch and K.W.~Becker, Eur. Phys. J. B {\bf 33},
  391 (2003).
\bibitem{Sykora} S. Sykora, A.~H\"ubsch, K.W.~Becker, G.~Wellein, and
  H.~Fehske, Phys. Rev. B {\bf 71}, 045112 (2005).
\bibitem{Hubsch_PAM} A. H\"{u}bsch and K.W.~Becker, Phys. Rev. B {\bf 71},
  155116 (2005).
\bibitem{spin} see, for example, G.S.~Uhrig, Phys. Rev. B {\bf 57}, R14004
  (1998); R.~Citro, E.~Orignac, and T.~Giamarchi, Phys. Rev. B {\bf 72},
  024434 (2005).
\bibitem{White} S.R.~White and I.~Affleck, Phys. Rev. B {\bf 54}, 9862
  (1996). 
\bibitem{Kosterlitz} J.M.~Kosterlitz and J.M.~Thouless, J. Phys. C {\bf 6},
  1181 (1973).
\bibitem{Caron} L.G.~Caron and S.~Moukouri, Phys. Rev. Lett. {\bf 76}, 4050
  (1996). 
\bibitem{Baxter} R.J.~Baxter, J. Phys. C {\bf 6}, L94 (1973).

\end{thebibliography}
\end{document}